\documentclass[twocolumn,english,journal]{IEEEtran}
\ifCLASSINFOpdf
\else
\fi
\usepackage[T1]{fontenc}
\usepackage{babel}
\usepackage{array}
\usepackage{calc}
\usepackage{amsmath}
\usepackage{amssymb}
\usepackage{graphicx}
\usepackage{cite}
\usepackage{algorithm}
\usepackage{subfigure}
\usepackage[unicode=true,
bookmarks=true,bookmarksnumbered=true,bookmarksopen=true,bookmarksopenlevel=1,
breaklinks=false,pdfborder={0 0 0},backref=false,colorlinks=false]
{hyperref}
\hypersetup{pdftitle={Your Title},
	pdfauthor={Your Name},
	pdfpagelayout=OneColumn, pdfnewwindow=true, pdfstartview=XYZ, plainpages=false}

\makeatletter

\providecommand{\tabularnewline}{\\}

\let\oldforeign@language\foreign@language
\DeclareRobustCommand{\foreign@language}[1]{%
	\lowercase{\oldforeign@language{#1}}}

\makeatother
\allowdisplaybreaks

\begin{document}
\title{Implication of Unobservable State-and-topology Cyber-physical Attacks}

\author{Jiazi~Zhang,~\IEEEmembership{Student Member,~IEEE,}
        Lalitha~Sankar,~\IEEEmembership{Senior Member,~IEEE}
        % <-this % stops a space
\thanks{J. Zhang and L. Sankar are with the Department
of Electrical and Computer Engineering, Arizona State University, Tempe,
AZ, 85281 USA e-mail: jzhan188@asu.edu, lalithasankar@asu.edu}}% <-this % stops a space
\maketitle

%\vspace{-0.3cm}
\begin{abstract}
%\boldmath
This paper studies the physical consequences of a
class of unobservable state-and-topology cyber-physical attacks in which both state and
topology data for a sub-network of the network are changed by an
attacker to mask a physical attack. The problem is formulated as a two-stage optimization
problem which aims to cause overload in a line of the network with limited attack resources.
It is shown that unobservable state-and-topology cyber-physical attacks as studied in this
paper can make the system operation more vulnerable to line
outages and failures.
\end{abstract}


% \fi
\IEEEpeerreviewmaketitle

\global\long\def\figurename{Fig.}
\global\long\def\tablename{TABLE}
%\vspace{-0.1cm}
\section{Introduction}
\IEEEPARstart{T}{he} electric power system is a complex cyber-physical system and is monitored by an intelligent which includes: (i) a supervisory control and data acquisition (SCADA) system; and (ii) an energy management system (EMS) that process the SCADA data. Network topology is important system data used in various
data processing modules in the EMS. Changes in topology can result from either system incidents or malicious physical attacks; but, in general, such topology alterations can be detected in the cyber layer. However, a sophisticated attacker can launch cyber attacks that alter the topology information in an unobservable manner; furthermore, they can also mask a physical attack via a cyber attack to create a more coordinated attack. Such cyber-physical attacks can result in wrong EMS solutions with potential serious consequences. Therefore, it is instructive to fully understand such attack consequences as a first step to thwart them.

There has been much recent interest in understanding both the physical and
cyber security challenges facing the electric power system. While there has been focus on the consequences of physical attacks on the system operation (\textit{e.g.,} \cite{Salmeron2004}), those of cyber as well as coordinated cyber-physical attacks are less understood. In this paper, we introduce a class of unobservable state-and-topology cyber-physical attacks in AC state estimation (SE) and focus on fully understanding its consequences. 

%We distinguish between \textit{cyber topology attack} and \textit{cyber-physical topology attack} implying the former only involves changes to measurement data (\textit{i.e.} cyber attack) while the later involves both a cyber attack and a physical attack.  
  
\vspace{-0.1cm}
\subsection{State of Art}

%There has been a lot of works on false data injection (FDI) attacks starting from DC attacks (\textit{e.g.},\cite{Liu2009}), effective attacks on AC SE (\textit{e.g.},\cite{Hug2012}), and consequences of attacks on electricity markets (\textit{e.g.}, \cite{Jia2014,Xie2011}) and system operations (\textit{e.g.}, \cite{Liang2014}).
%\textit{False data injection (FDI) attacks}: There has been a lot of works on false data injection (FDI) attacks starting from DC attacks (\textit{e.g.},\cite{Liu2009}), effective attacks on AC SE (\textit{e.g.},\cite{Hug2012}), and consequences of attacks on electricity markets (\textit{e.g.}, \cite{Jia2014,Xie2011}) and system operations (\textit{e.g.}, \cite{Liang2014}).  In \cite{Hug2012}, Hug and Giampapa 
%demonstrate that though AC SE is vulnerable to unobservable
%FDI attacks, it requires the knowledge of both system topology
%and states to launch such attacks.
\textit{False data injection (FDI) attacks}: In \cite{Liu2009}, Liu \textit{et al.} first introduce a class of FDI attacks on DC SE. In \cite{Hug2012}, Hug and Giampapa focus on
FDI attacks on AC SE and introduce a class of unobservable
attacks that are limited to a sub-graph of the networks. They
demonstrate that though AC SE is vulnerable to unobservable
FDI attacks, it requires the knowledge of both system topology
and states to launch such attacks. More recently, the attacks in \cite{Liang2014} by Liang \textit{et al.} study attack consequences by introducing
a class of unobservable FDI attacks for AC SE and demonstrate that such attacks can lead to a
physical generation re-dispatch and line overflow. 

% and show that an attacker with sufficient system
% knowledge can inject malicious measurements without being
% detected by existing bad data detection techniques that are
% subsequent to SE

\textit{Topology attacks}: Unobservable cyber attacks on topology can be of two types:\textit{
	line-maintaining} and\textit{ line-removing}. For a line-maintaining
attack, the attacker changes measurements and line status information
to make it appear that line that is not in the system is now shown
as active at the control center via SCADA data; the opposite is achieved
by a line-removing attack. For both line-removing and line-maintaining attacks, an attack can either change only topology data (\textit{i.e.,} state-preserving topology attack) or both state and topology data (\textit{i.e.}, state-and-topology attack). The class of unobservable cyber topology attacks is first introduced in \cite{Kim2013}; however, the analysis in \cite{Kim2013} is restricted to a subclass of
state-preserving line-removing attacks in which an
attacker changes topology information of the system without changing
the states. 
%In addition, the authors evaluate the attack's impact only on the electricity market. 

%Line-removing attacks are pure cyber attacks in which only cyber topology data is altered by attacker.
\textit{Line-maintaining attacks}: This sub-class of topology attacks require both physical line outage and cyber attack to mask the physical topology alteration and have not been studied yet in the literature. In this work, we study the line-maintaining cyber-physical attacks in which both physical and cyber topology are changed by the attacker. In \cite{Jzhang2015}, we consider unobservable state-preserving line-maintaining attacks (\textit{i.e.}, only topology data is changed) for which we develop an algorithm using breadth-first search (BFS) to find the smallest sub-network required to launch such an attack. However, changing only topology and not changing states limits the number of feasible lines amenable to attacks and also requires large load shifts at the end buses of a target line. Therefore, in this work, we determine attacks that change both state and topology.

\textit{Attack consequences}: There has been much focus on effect of attacks on operation costs \cite{Yuan11,Yuan2012} and electricity markets \cite{Xie2011,Jia2014}; in contrast, as in \cite{Liang2014, Liang2015}, this paper highlights physical system consequences of cyber-physical attacks. For cyber attacks whose goal is to effect electricity market and physical consequences,  
the attacks can be modeled as two-stage optimization problems where the first stage models the attack design with constraints that capture attacker's limitation and the second stages models the system response (see \cite{Yuan11, Yuan2012,Liang2015}). In this paper, we also use a two-stage optimization problem to find unobservable state-and-topology cyber-physical attack that can maximize power flow on a chosen line. 
Furthermore, due to the combination of physical and cyber attacks, we employ such a two-stage attack twice as detailed in the sequel.  

%\vspace{-0.2cm}
\subsection{Contributions}
The contributions of this paper are two-fold. First, we introduce a class of unobservable state-and-topology cyber-physical attacks in which
an attacker can change both cyber state and topology data to enable a coordinated physical and cyber attack on AC SE. Such an attack consists of a physical attack to first trip a transmission line, followed by a cyber attack that  
masks the physical attack. The goal is to overload a chosen line (different from tripped line) while avoiding being detected by both SE and the subsequent modules. 

Our attack model also captures the realistic limitation that the attacker can only access a sub-network of the entire power system, and therefore, can take down a line and modify the measurements only inside the sub-network. To this end, we can solve a two-stage optimization problem to determine the attack. However, since both physical attack and the re-dispatch resulting from cyber attack can lead to state changes, two attack vectors are required to enable the above two state changes and ensure the unobservability of the attack. Therefore, we formulate a two-step strategy to determine the attack. 
%In the first step, we solve a two-stage optimization problem to determine the transmission line to trip and an cyber attack vector for the attack sub-network after the system re-dispatches. In the second step, we determine an cyber attack vector for the system immediately after the physical attack that ensure the same dispatch. Note that the attacker only needs to solve the second step online; the first step can be solved offline since it does not involve knowledge of real-time system states. Such an attack strategy allows the attacker to change the SCADA data in a sub-network much faster in real-time.

The second contribution of our work is to demonstrate the consequences of the worst cyber-physical attacks determined by the proposed attack strategy on AC SE and AC OPF. We show that the cyber-physical attacks introduced here can successfully lead to line overflows in the IEEE 24-bus system with limited size of attack sub-network and load shifts.

The remainder of this paper is organized as follows. Sec.
\ref{Sec:Mathematical} introduces the general system model. Sec. \ref{sec:Gen-attack-model}
introduces the attack model.
Sec. \ref{sec:Worst-attack-strategy} presents a two-step attack strategy to
identify the worst-case overflow attack. Sec. \ref{sec:Simulation}
analyzes the numerical results for a test system. Sec. \ref{sec:conclusion} draws
the conclusion of this paper and presents the future
works. 

%\vspace{-0.1cm}
\section{\label{Sec:Mathematical}System Model}

\begin{figure}[h]
	\centering{}\includegraphics[trim=0.25cm 0.5cm 0.25cm 0.5cm, scale=0.46]{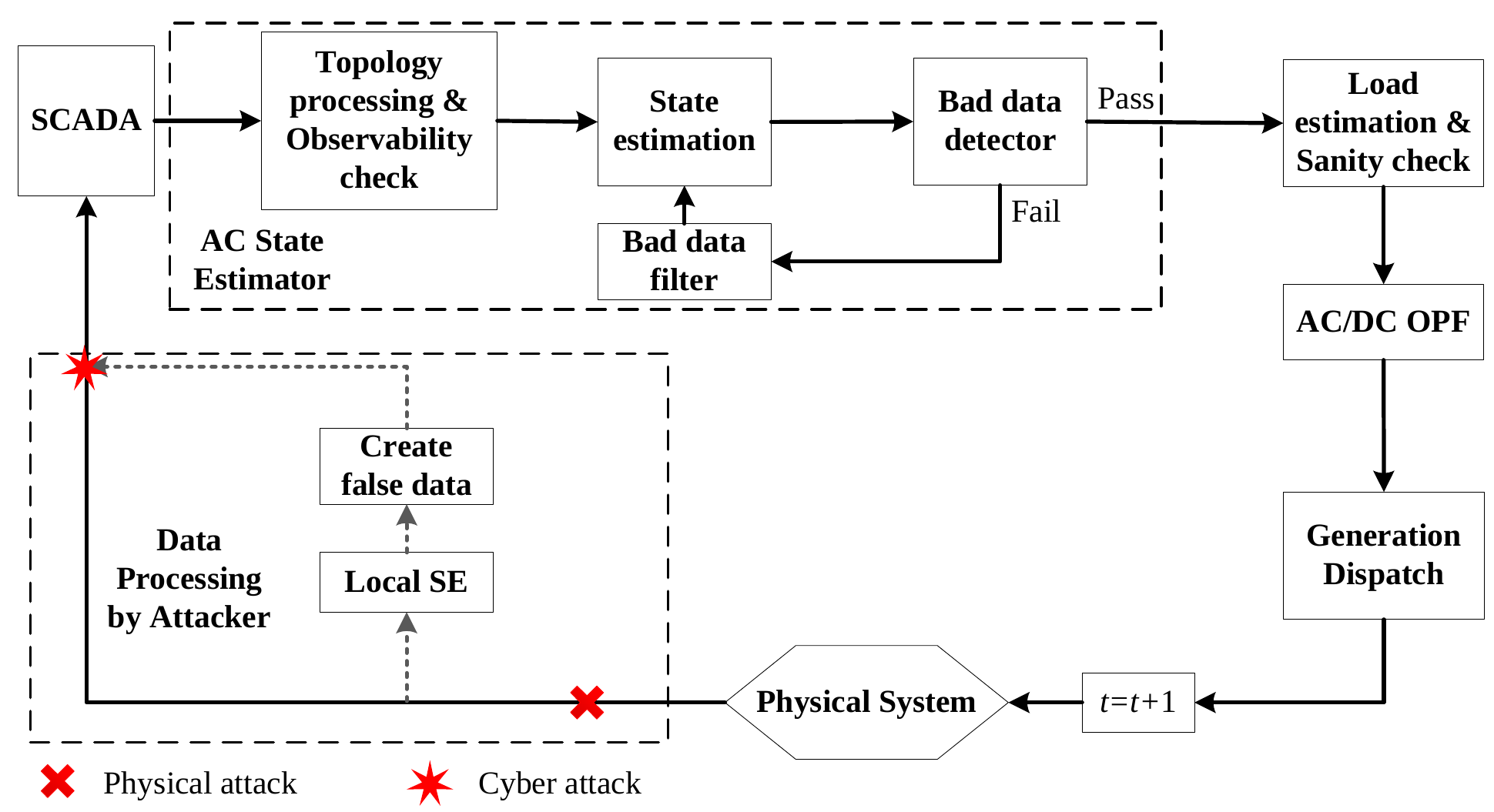}
	\protect\caption{Temporal Sequence of Data Processing Units in The Cyber Layer within Attack.\label{fig:Temporal-nature-of-1}}	
	%\vspace{-0.1cm}
\end{figure}

In this section, we introduce the mathematical formulation for the
various computational units of power system operation, including system network and topology, state estimation, and optimal power flow. Throughout,
we assume there are $n_{b}$ buses, $n_{br}$ branches, $n_{g}$ generators,
and $n_{z}$ measurements in the system. In Fig. \ref{fig:Temporal-nature-of-1}, we illustrate a typical temporal sequence of data processing units in the cyber layer.

%\vspace{-0.2cm}
\subsection{System Network and Topology}

The electric power system can be represented by a graph $\mathcal{G}=\left\{ \mathcal{N},\mathcal{E}\right\} $
where $\mathcal{N}$ and $\mathcal{E}$ are the sets of buses and
lines, respectively. 

At the control center, SCADA collects line status data as a $n_{br}\times1$
vector $s$ with entries $s_{k}\in\left\{ 0,1\right\} $ for $k\in\{1,...,n_{br}\}$
that indicate the on and off status of circuit-breakers on each line.
The data is then passed to a topology processor to map the real-time
power system topology. Each $s$ corresponds to a specific system topology $\mathcal{G}$. 

%\vspace{-0.4cm}
\subsection{State Estimation}

Consider an $n_{z}\times1$ vector $z$ of nonlinear measurements (for AC SE)
given as 
\begin{equation}
z=h(x,\mathcal{G})+e\label{eq:ACMeasurement}
\end{equation}
where $x=\left[\theta,\: V\right]^{T}$ is the system state vector,
and $e$ is an $n_{z}\times1$ noise vector which is independent of
$x$ and is modeled as Gaussian distributed with $0$ mean and $\sigma_{i}^{2}$
covariance such that the measurement error covariance matrix is given
by $R=diag(\left\{ \sigma_{i}^{2}\right\} _{i=1}^{M})$. The function
$h(x,\mathcal{G})$ is a vector of nonlinear functions that describes
the relationship between the system states and measurements for a
topology $\mathcal{G}$. 

We use weighted least-squares (WLS) AC SE to calculate the $\theta$
and $V$ \cite{AburBook}. Subsequent to SE, bad data detector use $\chi^{2}-$test to detect bad
data and bad data identification use largest normalized
residual method to filter the bad data. 
 %The states are solved as a least
%square problem with the following objective function 
%\begin{equation}
%\min J(x)=(h(x)-z)^{T}R^{-1}(h(x)-z),\label{eq: problem (2)}
%\end{equation}
%the solution to which satisfies 
%\begin{equation}
%g(\hat{x})=\frac{\partial J(\hat{x})}{\partial x}=H^{T}(\hat{x})\cdot R^{-1}\cdot(h(\hat{x})-z)=0\label{eq: GainMatrix}
%\end{equation}
%where the system Jacobian matrix $H=\frac{\partial h(x)}{\partial x}\mid_{x=\hat{x}}$,
%and $\hat{x}$ is the $2n_{b}\times1$ estimated state vectors. The
%WLS solution for this nonlinear optimization problem can be solved
%iteratively.

%\vspace{-0.3cm}
\subsection{Optimal Power Flow}

The OPF problem can be written as 
\begin{flalign}
\text{min}\;\; & C\left(x\right)\label{eq:OPF_obj}\\
s.t.\;\; & F\left(x\right)=0\label{eq:OPF_cons1}\\
& T\left(x\right)\leqslant0\label{eq:OPF_cons2}\\
& x_{\text{min}}\leqslant x\leqslant x_{\text{max}}\label{eq:VariableLimit}
\end{flalign}
where $x=\left[\theta,V,P_{G},Q_{G}\right]^{T}$is the optimization
vector with voltage angle $\theta$, voltage magnitude $V$ that are
both $n_{b}\times1$ vectors, and active power generation $P_{G}$,
reactive power generation $Q_{G}$ that are both $n_{g}\times1$ vectors;
$C(\cdot)$ denotes the cost function of $x$; $F\left(\cdot\right)$
denotes the equivalent constraints (power balance constraints); $T\left(\cdot\right)$ denotes
the inequivalent constraints (power flow limits).

\vspace{-0.2cm}
\section{\label{sec:Gen-attack-model}Attack Model}
The unobservable state-and-topology cyber-physical attack considered here models both a physical attack and a coordinated cyber attack. 

%\subsection{aaa}
We assume the attacker has the following capabilities: 
\begin{enumerate}
	\item Attacker has knowledge of the topology $\bar{\mathcal{G}}_{0}$
	of entire network prior to physical attacks. 
	\item Attacker has the capability to launch physical attack, and observe and
	change measurements only for a sub-graph $\mathcal{S}$ of
	$\mathcal{G}_{0}$. The choice of $\mathcal{S}$ is described in
	detail in the sequel.
	\item Attacker has the capability to perform SE and compute modified measurements
	for $\mathcal{S}$. 
	\item Attacker has knowledge of the capacity and operation \\ cost of every generator in the network.
	\item Attacker has historic data of load patterns and generation dispatch of the entire network.
\end{enumerate}
We assume that the power system is observable before and after the physical attack. 

In this paper, we focus only on physical attacks that target transmission lines. We denote the line that is physically tripped by the attacker as the \textit{switching attack line} and the two end buses of this line as the \emph{switching attack buses}. Assume the switching attack line is line $t$ and the topology prior to the physical attack is $\mathcal{G}_{0}$. The physical line status for line $t$ changes from $s_{t}=1$ to $s_{t}=0$ after the physical attack and the corresponding physical topology changes to $\mathcal{G}$.

In general, a physical attack will be subsequently detected by the topology processing unit in the EMS and the system topology will be updated shortly after the detection. However, a sophisticated attacker can hide such physical attacks by launching an unobservable cyber attack. In the resulting \textit{unobservable cyber topology attack}, the attacker modifies line status as
well as related bus measurements to alter the system topology $\mathcal{G}$
to a different ``target'' topology $\bar{\mathcal{G}}\negthinspace=\negthinspace\{\mathcal{N}\negthinspace,\bar{\mathcal{E}}\}$.
Since the attacker's aim is to hide the topology alteration caused by the physical attack, $\bar{\mathcal{G}}$ should be chosen as $\mathcal{G}_{0}$.

To launch a state-and-topology attack, the attacker injects
$n_{br}\times1$ line status attack vector $b$ and $n_{z}\times1$
measurement attack vector $a$. The attack vector $b$ for line status overrides the physical change on line $t$'s status by setting for $b_{k}\negthinspace=\negthinspace 0$ for $k\neq t$ and $b_{k}\negthinspace=\negthinspace1$ for $k\negthinspace=\negthinspace t$.
These changes lead
to a new system state $\bar{x}$ for the system under attack. This
attack modifies $\left(s,z\right)$ for topology $\mathcal{G}$ to
$\left(\bar{s},\bar{z}\right)$ for topology $\bar{\mathcal{G}}$
such that
\begin{equation}
\bar{s}=s+b,\; \; \textmd{and}\; \bar{z}=z+a.\label{eq: ModMeasurements}
\end{equation}
In the absence of noise, the measurement attack vector satisfies
\begin{equation}
a=h(\bar{x},\bar{\mathcal{G}})-h\left(x,\mathcal{G}\right).\label{eq: AttackVector}
\end{equation}

%The simplest subclass of unobservable topology attacks that can hide
%the physical attacks are state-preserving attacks as introduced in \cite{Kim2013} and \cite{Jzhang2015}. These attacks aim
%to change the system topology information while preserving the states.
%Attacker only need to modify (a) the line status data of switching
%attack line, (b) power flow measurements on the line, and (c) the power
%injection measurements on buses connecting to the switching attack
%line to accomplish such attacks. Therefore, the resources required
%by state-preserving topology attacks are relatively less. 
%However, since the power injection measurements can only be altered
%by shifting loads in cyber layer, attackers are limited to a small
%set of switching attack lines which connecting two load buses. Another crucial limitation of these
%attacks is that the load shifts caused by such attacks can be greater
%than 10\% on load buses. Such
%a change can lead to detection of anomaly. 

%\begin{figure}[h]
%	\centering\includegraphics[trim=0.55cm 1cm 0.7cm 1cm, clip=true, scale=0.75]{CompromiseGraph2}\protect
%	\protect\caption{Example of Modified Meters and Line Status Data for (a) Unobservable State-preserving Topology Attack; and (b) Unobservable Joint Topology and State Attack.\label{fig:An-example-of}}
%\end{figure}

%To avoid these defects, we focus on state-and-topology attacks, in which
%attacker change both states and topology information, hereinbelow. 

For nonlinear measurement model and AC SE,  we model a sophisticated attacker who attacks measurements
and line status data for a sub-graph $\mathcal{S}$ of the network
by first estimating the system states $\hat{x}$ inside
$\mathcal{S}$ using AC SE. The attacker then chooses a small set
of buses in $\mathcal{S}$ to change states from the estimate $\hat{x}$
to $\bar{x}=\hat{x}+c$ such that the
measurement vector $\bar{z}$ after cyber attack has entries 
%\vspace{-0.1cm}
\begin{align} 
\bar{z}_{i} & =\begin{cases}
\begin{array}{l}
z_{i}\:,\\
h_{i}(\hat{x}+c,\mathcal{\bar{G}})\:,
\end{array} & \begin{array}{l}
i\notin\mathcal{I}_{\mathcal{S}}\\
i\in\mathcal{I}_{\mathcal{S}}
\end{array}\end{cases}.\label{eq: ModMeasureStra1} 
\end{align}
where $\mathcal{I}_{\mathcal{S}}$ denotes the set of measurements inside $\mathcal{S}$. 

We use the following method to identify the sub-graph $\mathcal{S}$ for an unobservable state-and-topology attack. Throughout, we distinguish two types of buses: \textit{load buses} with presence of load and \textit{non-load buses} with no load.
\begin{enumerate}
	\item Use the optimization problem (the details are in the sequel) to determine the load buses from the attack vector $c$ whose states need to be changed (defined as \textit{center bus}) to enable the attack. 
	\item Include all center
	buses in $\mathcal{S}$. 
	\item Extend $\mathcal{S}$ by including
	all buses and branches connected to the buses inside $\mathcal{S}$. 
	\item If there are non-load buses on the boundary of $\mathcal{S}$,
	extend $\mathcal{S}$ by including all adjacent buses of the non-load
	boundary buses and the corresponding branches.
	\item Repeat 4) until all boundary buses of $\mathcal{S}$
	are load buses. 
	\item Check if there is a path (actual bus and branch connection) in $\mathcal{S}$ that can
	connect the two switching attack buses. If such path
	exits, then $\mathcal{S}$ is the attack sub-graph. If there is no
	such path, go to Step 7). 
	\item Use BFS method to find the shortest path connecting the two
	switching attack buses. Include the shortest path in
	$\mathcal{S}$. Then this $\mathcal{S}$ is the attack sub-graph.
\end{enumerate}

%We explain the reasons for the steps as follows.

Steps 1)$-$5) ensure the boundary buses of $\mathcal{S}$ are load buses with states unchanged. For a non-load bus in $\mathcal{S}$, since the injection of non-load buses are known to the control center,
the attacker should ensure that under an attack, the net injection is equal to the net flow into the bus. Thus, the state changes for non-load buses are dependent on those for the neighboring load buses. Furthermore, the state of a boundary bus $j$ is computed using both measurements inside and outside $\mathcal{S}$. From \eqref{eq: ModMeasureStra1}, if a measurement $i$ for $i\notin\mathcal{I}_{\mathcal{S}}$ is dependent on the $j^{\text{th}}$ state, then the corresponding $j^{\text{th}}$ entry of the attack vector should satisfy $c_{j}=0$ to ensure the attack to be unobservable. Thus, a boundary bus cannot have a state change, and therefore, cannot be a non-load buses.

Steps 6) and 7) ensure that the states of switching attack buses can be estimated with measurements inside $\mathcal{S}$. To maintain the switching attack line as active in the cyber layer, the attacker needs to modify the line status as well as power flow measurements on the switching attack line and power injection measurements on the switching attack buses. This in turn, requires the attacker to estimate the states of switching attack buses to create the false measurements. However, since this line is physically disconnected, the attacker needs to use an algorithm such as BFS to determine an alternate shortest path connecting the 2 switching attack buses, and thereby estimate the states and changed measurements. In general, state change is required for at least one of the switching attack buses. This bus, thereby, will be included in $\mathcal{S}$. However, $\mathcal{S}$ may not include the entire physical path. Thus, the attacker needs steps 6) and 7) to complete the path.

%Sub-graph identified with this method ensures that only loads inside it are changed by attacker while the net load of the whole system remain the same.
%It also allows the states
%of the boundary buses to remain unchanged via changes to load injections.
%Since large changes in load estimates can be detected, attacker
%has to determine the state change vector $c$ such that the load shifts
%at all load buses in $\mathcal{S}$ are bounded. 

%\vspace{-0.4cm}
\section{\label{sec:Worst-attack-strategy}Attack Strategy }

In this section, we study the worst-case cyber-physical attacks. We assume the attacks can: (a) physically trip a switching attack line and mask the physical attack with a cyber attack; (b) maximize power flow on a \textit{target line}; and (c) avoid detectability by limiting load shift via changes in measurements. The attack resources available to the attacker may also be limited. We model this limitation by constraining the size of sub-network the attacker has access to. This leads to a constrained optimization problem. As noted before, two attack vectors are needed for the physical and cyber parts of the attack and each optimization problem is described below.

Our two-step optimization problem captures the temporal nature of attack sequence involving a physical attack followed by several cyber attacks.  

\begin{figure}[h]
	\centering{}\includegraphics[scale=0.45]{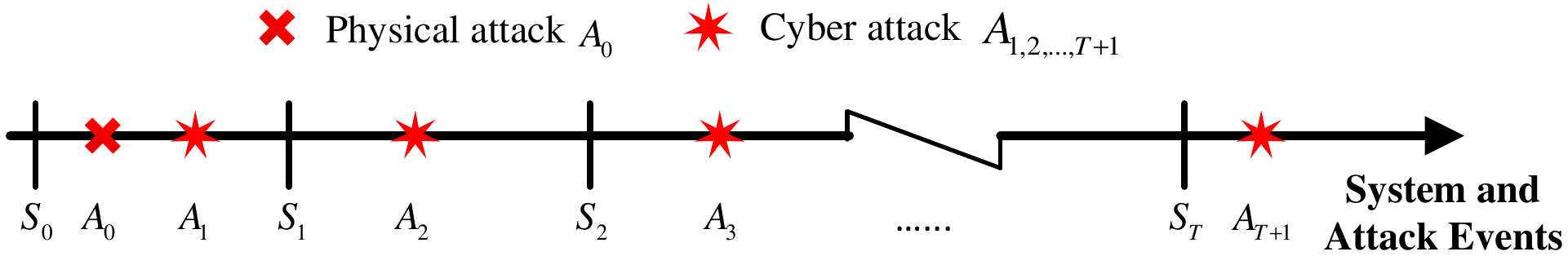}\protect\caption{Time sequence of attack and system events.\label{fig:worstAttack_TS}} 
	\vspace{-0.2cm}
\end{figure}

In Fig. \ref{fig:worstAttack_TS}, we illustrate this temporal sequence of attack and system events. The system events are periodic and are denoted by $S_{t}$ for the $t^{\text{th}}$ event. At the start of each $S_{t}$, data is collected from SCADA and by the end of $S_{t}$, \textit{i.e.,} the start of $S_{t+1}$, data is processed in the EMS. There are 2 attacks instance, $A_{0}$ and $A_{1}$ to denote the physical and cyber attack events, respectively. We assume the physical attack event $A_{0}$ is launched immediately after the start of the $0^{\text{th}}$ system event, \textit{i.e.,} $S_{0}$, and the coordinated cyber attack event $A_{1}$ is launched shortly after, but before the start of next system event $S_{1}$. Following this cyber-physical attack pair ($A_{0}$, $A_{1}$), the cyber attack is sustained between every two system events to maintain the worst generation dispatch, and thereby, sustain the maximal power flow on the target line. In TABLE \ref{tab:PH_CY_DATA}, we denote how the cyber (measured) and physical (actual) data including generation dispatch, system state, topology, and loads vary at all system and attack events.
%\vspace{-0.2cm}

\begin{table}[h]
	\protect\protect\caption{Physical and cyber data for attack and system events.\label{tab:PH_CY_DATA}}
	
	\centering{}%
	\begin{tabular}{|>{\centering}m{0.8cm}|>{\centering}m{0.2cm}|>{\centering}m{0.2cm}|>{\centering}m{0.4cm}|>{\centering}m{0.35cm}|>{\centering}m{0.35cm}|>{\centering}m{0.35cm}|>{\centering}m{0.35cm}|>{\centering}m{0.2cm}|>{\centering}m{0.35cm}|>{\centering}m{0.35cm}|}
		\hline 
		$\negthickspace\negthinspace$ System and Attack Event  & $E_{0}$  & $A_{0}$  & $A_{1}$ & $S_{1}$ & $A_{2}$ & $S_{2}$ & $A_{3}$ & ... & $S_{T}$ & $\negthinspace A_{T\negthinspace+\negthinspace1}$ \tabularnewline
		\hline 
		$\negthickspace\negthinspace$Generation $\negthickspace$Dispatch & $P_{G}^{0}$ & $P_{G}^{0}$ & $P_{G}^{0}$ & $P_{G}^{*}$ & $P_{G}^{*}$ & $P_{G}^{*}$ & $P_{G}^{*}$ & ... & $P_{G}^{*}$ & $P_{G}^{*}$\tabularnewline
		\hline 
		Physical $\negthinspace\negthinspace\negthickspace$Topology & $\bar{\mathcal{G}}$ & $\mathcal{G}$ & $\mathcal{G}$ & $\mathcal{G}$ & $\mathcal{G}$ & $\mathcal{G}$ & $\mathcal{G}$ & ... & $\mathcal{G}$ & $\mathcal{G}$\tabularnewline
		\hline 
		Cyber $\negthinspace\negthinspace\negthickspace$Topology & $\bar{\mathcal{G}}$ & $\mathcal{Gq	1}$ & $\bar{\mathcal{G}}$ & $\bar{\mathcal{G}}$ & $\bar{\mathcal{G}}$ & $\bar{\mathcal{G}}$ & $\bar{\mathcal{G}}$ & ... & $\bar{\mathcal{G}}$ & $\bar{\mathcal{G}}$\tabularnewline
		\hline 
		Physical State & $\theta_{0_{-}}$ & $\theta_{0}$ & $\theta_{0}$ & $\theta^{*}$ & $\theta^{*}$ & $\theta^{*}$ & $\theta^{*}$ & ... & $\theta^{*}$ & $\theta^{*}$ \tabularnewline
		\hline 
		Cyber State & $\theta_{0_{-}}$ & $\theta_{0}$ & $\negthinspace\negthinspace\negthickspace\theta_{0}\negthickspace+\negmedspace c^{0}\negthickspace\negthinspace\negthinspace$ &  $\negthinspace\negthinspace\negthickspace\theta^{*}\negthickspace+\negmedspace c\negthickspace\negthinspace\negthinspace$ &  $\negthinspace\negthinspace\negthickspace\theta^{*}\negthickspace+\negmedspace c\negthickspace\negthinspace\negthinspace$ &  $\negthinspace\negthinspace\negthickspace\theta^{*}\negthickspace+\negmedspace c\negthickspace\negthinspace\negthinspace$ &  $\negthinspace\negthinspace\negthickspace\theta^{*}\negthickspace+\negmedspace c\negthickspace\negthinspace\negthinspace$ & ... &  $\negthinspace\negthinspace\negthickspace\theta^{*}\negthickspace+\negmedspace c\negthickspace\negthinspace\negthinspace$ &  $\negthinspace\negthinspace\negthickspace\theta^{*}\negthickspace+\negmedspace c\negthickspace\negthinspace\negthinspace$ \tabularnewline
		\hline 
		Physical Load & $P_{D}$ & $P_{D}$ & $P_{D}$ & $P_{D}$ & $P_{D}$ & $P_{D}$ & $P_{D}$ & ... & $P_{D}$ & $P_{D}$\tabularnewline
		\hline 
		Cyber Load & $P_{D}$ & $P_{D}$ & $\bar{P}_{D}$ & $\bar{P}_{D}$ & $\bar{P}_{D}$ & $\bar{P}_{D}$ & $\bar{P}_{D}$ & ... & $\bar{P}_{D}$ & $\bar{P}_{D}$ \tabularnewline
		\hline 
	\end{tabular}
	\vspace{-0.1cm}
\end{table}	

Assume the system topology and the generation at $S_{0}$ are $\bar{\mathcal{G}}$ and $P_{G}^{0}$, respectively. From TABLE \ref{tab:PH_CY_DATA}, we can see that the system physical topology changes to $\mathcal{G}$ after the physical attack. The physical operation states, thereby, change to $\theta_{0}$. The attacker then injects cyber attack vector $c^{0}$ to change the load pattern from the physical load $P_{D}$ to the false cyber load $\bar{P}_{D}$ to mask the physical topology alteration. The physical and cyber loads at attack event $A_{1}$ satisfy the following relationships, respectively: 
\begin{equation}
P_{D}\negthinspace=\negthinspace A_{GN}P_{G}^{0}\negthinspace-\negthinspace H_{1}\theta_{0},\; \text{and} \; \bar{P}_{D} =A_{GN}P_{G}^{0}\negthinspace-\negthinspace\bar{H}_{1}\left(\negthinspace\theta_{0}\negthinspace+\negthinspace c^{0}\right)\label{eq:Pd_cyber}
\end{equation}
where $A_{GN}$ is $n_{b}\times n_{g}$ generator-to-bus connectivity matrix; $H_{1}$ and $\bar{H}_{1}$ are $n_{b}\times n_{b}$ dependency matrices
between power injection and voltage angle for $\mathcal{G}$ and $\bar{\mathcal{G}}$, respectively. When subtracting the two equations in \eqref{eq:Pd_cyber}, the cyber loads are related to the physical loads as
\begin{equation}
\bar{P}_{D}=P_{D}+H_{1}\theta_{0}-\bar{H}_{1}(\theta_{0} + c^{0})  \label{eq:falseload_A1}
\end{equation}
The false cyber load $\bar{P}_{D}$ and topology $\bar{\mathcal{G}}$ leads to a system re-dispatch to the optimal generation dispatch $P_{G}^{*}$ at $S_{1}$. Since the attacker optimization problem at each step models the system response, such an optimal dispatch will cause maximal power flow on the target line. Following this first cyber attack $A_{1}$, since the generation dispatch changes at $S_{1}$, the physical system states also change to $\theta^{*}$. To sustain both the optimal dispatch $P_{G}^{*}$ and the false cyber topology $\bar{\mathcal{G}}$ at the next system event, \textit{i.e.,} $S_{2}$, the attacker needs to maintain the false cyber load $\bar{P}_{D}$ by injecting another attack vector $c$ at $A_{2}$. Thus, the nodal power balance at attack event $A_{2}$ in the cyber layer is:
 \begin{equation}
 A_{GN}P_{G}^{*}-\bar{H}_{1}\left(\theta^{*}+c\right)=P_{D}+H_{1}\theta_{0}-\bar{H}_{1}\left(\theta_{0}+c^{0}\right).\label{eq:NB_CONS}
 \end{equation}
 where the right hand side terms represent the cyber load modified at $A_{1}$. 
In the following attack events, \textit{i.e.,} $A_{t}, \; t=3,...,T$, the attacker can keep injecting $c$ to maintain the false cyber load $\bar{P}_{D}$. This in turn ensures that the optimal dispatch and the false cyber topology are maintained at $P_{G}^{*}$ and $\bar{\mathcal{G}}$, respectively, and the maximal power flow on the target line is sustained.
          
To model the cyber-physical attack events $A_{0}$, $A_{1}$, and $A_{2}$ between $S_{0}$ and $S_{1}$, the optimization problem should capture the power balance relationship shown in \eqref{eq:NB_CONS}. However, since the switching attack line is determined by the optimization problem, both $H_{1}$ and $\theta_{0}$ are unknown before solving the problem. On the other hand, for the pure cyber attack events $A_{2}$ and $A_{3}$, the power balance in the cyber layer is
\begin{equation}
A_{GN}P_{G}^{*}-\bar{H}_{1}\left(\theta^{*}+c\right)=P_{D}+H_{1}\theta^{*}-\bar{H}_{1}\left(\theta^{*}+c\right).\label{eq:NB_CONS2}
\end{equation} 
This is equivalent to the physical power balance as
\begin{equation}
A_{GN}P_{G}^{*}-H_{1}\theta^{*}=P_{D}.\label{eq:NB_CONS1}
\end{equation}

 Therefore, instead of directly modeling the cyber-physical attack events $A_{0}, A_{1}$, and $A_{2}$ between $S_{0}$ and $S_{1}$, we can model the pure cyber attack events $A_{2}$ and $A_{3}$ between $S_{1}$ and $S_{2}$ to determine the attack vector $c$ in the first step. Such a $c$ should be subject to bounds on both the attacker's sub-graph size and the load shifts.
However, since the new topology $H_{1}$ is still not known prior to the optimization, we replace $H_{1}$ using the following equations:
 \begin{flalign}
 H_{1}\theta^{*} & =A_{KN}P_{K}\label{eq:PF1}\\
 P_{K} & =\text{diag}\left(s\right)\cdot\bar{H}_{2}\theta^{*}\label{eq:PF2}
 \end{flalign} 
where $A_{KN}$ is the $n_{b}\times n_{br}$ branch-to-bus connectivity matrix, $\bar{H_{2}}$ is the $n_{br}\times n_{b}$ dependency matrix between
power flow and voltage angle for $\bar{\mathcal{G}}$, $s$ is the line status vector, $diag(s)$ represents the diagonal matrix of $s$, $P_{K}$ is the $n_{br}\times 1$ power flow vector. In \eqref{eq:PF1}, the sum of physical power flows on the set of branches connected to a bus is utilized to calculate the physical power injection at the bus. In \eqref{eq:PF2}, the physical power flow vector is represented by the diagonal matrix of line status vector $s$ multiply the cyber power flow vector, \textit{i.e.,} $\bar{H}_{2}\theta^{*}$. That is, if a line $t$ is selected as the switching attack line, the power flow $P_{Kt}$ on line $t$ is forced to be $0$, otherwise, $P_{Kt}=\bar{H}_{2}({t},:)\cdot\theta$, where $\bar{H}_{2}(t,:)$ represents the $t^{\text{th}}$ row of $\bar{H}_{2}$.    
 With these modifications, we can then model the system and cyber attack events from $A_{2}$ through $S_{2}$ to $A_{3}$ with a two-stage optimization problem, and hence, the switching attack line can also be determined as the solution of the optimization problem. After the switching attack line and the cyber attack vector $c$ are both determined, the attack sub-graph $\mathcal{S}$ can be identified with the process stated in Section \ref{sec:Gen-attack-model}. The details of this problem is described in Subsection \ref{sub:Maximize-power-flow}. 
 %This step can be assumed as an off-line analysis since no real-time information is required in the two-stage optimization problem. 

In the second step, we focus on the attack vector $c^{0}$ at $A_{1}$. We again use a two-stage optimization problem to determine the $c^{0}$ such that the optimal generation dispatch for this problem is forced to be same as that in Step 1. We, henceforth, define the attack vector solved in the second step as the \emph{initial attack vector}. The details of the second step is introduced in Subsection \ref{sub:Initial-attack-vector}. 
%This step can be assumed as an on-line attack determination since it requires the real-time physical states data. 

The attack vectors $c$ and $c^{0}$ are both DC attack vectors that can be detected by AC SE. Thus, to ensure the unobservability of the attacks, the attacker should construct two AC attacks with $c$ and $c^{0}$. This procedure is introduced in Subsection \ref{sub:Implementation}. 

%\vspace{-0.5cm}
\subsection{\label{sub:Maximize-power-flow}Step 1: Maximize Power Flow on A Line}

In Step 1, we introduce a two-stage optimization problem to determine the attack vector $c$ and
the switching attack lines such that the target line $l$ in the attacker's sub-graph
$\mathcal{S}$ has maximal power flow subject to specific constraints as explained in the sequel. The two-stage optimization
is given as
\begin{flalign}
\text{max}\:\; & P_{Kl}-\zeta\left\Vert c_{\mathcal{L}}\right\Vert _{0}\label{eq:Obj1_MaxPF}\\
\text{s.t.}\hspace{0.2cm}\;
& \begin{array}{cc}
\underset{k=1}{\overset{n_{br}}{\sum}}(1-s_{k})=N_{T}, & s_{k}\in\left\{ 0,1\right\} \end{array}\label{eq:con_tripline}\\
& \left\Vert c_{\mathcal{L}}\right\Vert _{0}\leqslant N_{0}\label{eq:con_resources}\\
& -\tau P_{D}\leqslant\bar{H}_{1}\left(\theta^{*}+c\right)-A_{KN}P_{K}^{*}\leqslant\tau P_{D}\label{eq:con_loadshift}\\
& \left\{ \theta^{*}, P_{G}^{*},P_{K}^{*}\right\} =arg\left\{ \underset{\theta,P_{G},P_{K}}{min}\:\underset{g=1}{\overset{n_{g}}{\sum}} C_{g}\left(P_{Gg}\right)\right\} \label{eq:OBJ_MINCOST}\\
& \text{s.t.}\hspace{0.1cm}A_{GN}P_{G}-A_{KN}P_{K}=P_{D}\:(\lambda)\label{eq:con_nodebalance}\\
& \hspace{0.4cm} P_{K}=\text{diag}\left(s\right)\cdot\bar{H}_{2}\theta\label{eq:con_truePF}\\
& \begin{array}{cc}
\hspace{0.2cm}- P_{K}^{\text{max}}\leqslant\bar{H}_{2}\left(\theta+c\right)\leqslant P_{K}^{\text{max}} & (\mu^{-},\mu^{+})\end{array}\label{eq:con_powerflow}\\
& \begin{array}{cc}
\ \, P_{G}^{\text{min}}\leqslant P_{G}\leqslant P_{G}^{\text{max}} & (\alpha^{-},\alpha^{+})\end{array}\label{eq:con_GENlimit}
\end{flalign}
where $C_{g}\left(\cdot\right)$ is the cost function for generator $g$; $P_{G}$
is $n_{g}\times1$ active power generation vector with maximum and
minimum limit $P_{G}^{\text{max}}$ and $P_{G}^{\text{min}}$ , respectively;
$P_{K}$ is $n_{br}\times1$ physical power flow vector with thermal
limit $P_{K}^{\text{max}}$; $\lambda$ is $n_{b}\times1$ dual variable
vector of constraint \eqref{eq:con_nodebalance}; $\mu^{\mp}$ are $n_{br}\times1$
dual variable vectors of constraint \eqref{eq:con_powerflow},
respectively; $\alpha^{\mp}$ are $n_{b}\times1$ dual variable vectors
of constraint \eqref{eq:con_GENlimit},
respectively; $\bar{H_{1}}$ is $n_{b}\times n_{b}$ dependency matrix
between power injection and voltage angle for $\mathcal{\bar{G}}$;
$\bar{H_{2}}$ is $n_{br}\times n_{b}$ dependency matrix between
power flow and voltage angle for $\mathcal{\bar{G}}$; $P_{D}$
is $n_{b}\times1$ physical load vector, which
has maximum load shift percentage $\tau$; $\zeta$ is the weight
of the norm of attack vector $c$; $\mathcal{L}$ represents the set of load buses; $N_{0}$ is the maximum number of load-buses that can be attacked; $N_{T}$ is the maximum number of switching attack lines.

The goal of the attack in \eqref{eq:Obj1_MaxPF} is a multi-objective problem which includes maximizing the
power flow on the target line $l$ to create an overflow, while minimizing
the $l_{0}$-norm of the attack vector, \textit{i.e.}, minimizing the attack sub-graph size. The power flow on $l$ is maximized along the direction
of the power flow prior to attack. In the first stage, constraints \eqref{eq:con_tripline}$-$\eqref{eq:con_loadshift} model the following attacker limitations: (i) only up to $N_{T}$ switching attack lines can be physically tripped; (ii) alter up to $N_{0}$ load-bus states; and (iii) limit cyber load shifts to at most $\tau P_{D}$; respectively. The second stage optimization represents DC OPF, whose aim is to minimize
operation cost in \eqref{eq:OBJ_MINCOST}, subject to power balance constraints in \eqref{eq:con_nodebalance} and \eqref{eq:con_truePF},
thermal limit constraint in \eqref{eq:con_powerflow}, and generation limit constraint in \eqref{eq:con_GENlimit}.

This two-stage optimization problem is nonlinear and non-convex. For
tractability, we modify several constraints.

Constraint \eqref{eq:con_truePF} is a nonlinear constraint which
includes the product of binary variable $s$ and continuous variable
$\theta$. It can be replaced by a linear form as follows 
\begin{equation}
\left\{ \begin{array}{lr}
-P_{K}+\bar{H}_{2}\theta^{*}\leqslant M_{1}\left(1-s\right) & \left(\beta^{-}\right)\\
P_{K}-\bar{H}_{2}\theta^{*}\leqslant M_{1}\left(1-s\right) & \left(\beta^{+}\right)\\
-P_{K}\leqslant M_{1}\cdot s & \left(\gamma^{-}\right)\\
P_{K}\leqslant M_{1}\cdot s & \left(\gamma^{+}\right)
\end{array}\right.\label{eq:con_linearPF}
\end{equation}
where $\beta^{\pm}$ and $\gamma^{\pm}$ are $n_{br}\times1$ dual
variable vectors for the corresponding constraints and $M_{1}$ is
a large number.

Constraint \eqref{eq:con_resources} is an $l_{0}-$norm constraint on
the attack vector, which is nonlinear
and non-convex. It can be relaxed to a corresponding
$l_{1}-$norm constraint as:
\begin{equation}
\left\Vert c_{\mathcal{L}}\right\Vert _{1}=\underset{^{n\in\mathcal{L}}}{\sum}\left|c_{n}\right|\leqslant N_{1}.\label{eq:NORM1}
\end{equation}

However, constraint \eqref{eq:NORM1} is still nonlinear. We, thus,
linearize it as follows:
\begin{equation}
c_{n}\leqslant u_{n}, \hspace{0.5cm}-c_{n}\leqslant u_{n}, \hspace{0.5cm}\underset{^{n\in\mathcal{L}}}{\sum}u_{n}\leqslant N_{1}.
\label{eq:con_linearN1}
\end{equation}
where $u$ is $n_{load}\times1$ non-negative slack variable vector. 

Once the attack vector determined by $s$ and $c$ is given
in the first stage optimization problem, the second stage DCOPF problem
\eqref{eq:OBJ_MINCOST}$-$\eqref{eq:con_GENlimit} and \eqref{eq:con_linearPF}
is then convex. The second stage optimization problem can then be
replaced by its Karush-Kuhn-Tucker (KKT) optimality conditions as follows:
\begin{flalign}
& \nabla\left(\underset{g=1}{\overset{n_{g}}{\sum}} C_{g}\left(  P_{Gg}^{*} \right) \right) + \lambda^{T} \cdot \nabla \left(  A_{GN}  P_{G}^{*} -  A_{KN}  P_{K}^{*} -  P_{D} \right)\vspace{-0.2cm}\nonumber \\
& +\left[\mu^{-};\mu^{+}\right]^{T} \!\!\cdot\nabla \left(\!\left[ \begin{array}{c}
\!\!-\bar{H}_{2}\left(\theta^{*}+c\right)\!\!\\
\!\!\bar{H}_{2}\left(\theta^{*}+c\right)\!\!
\end{array} \right] - \left[ \begin{array}{c}
\!\!P_{K}^{\text{max}}\!\!\\
\!\!P_{K}^{\text{max}}\!\!
\end{array} \right]\!\right)\nonumber \\
& +\left[\alpha^{-};\alpha^{+}\right]^{T} \!\!\cdot\nabla \left( \!\left[ \begin{array}{c}
\!\!-P_{G}^{*}\!\!\\
\!\!P_{G}^{*}\!\!
\end{array} \right] - \left[ \begin{array}{c}
\!\!-P_{G}^{\text{min}\!\!}\\
\!\!P_{G}^{\text{max}\!\!}
\end{array} \right] \right)\nonumber \\
& +\left[\beta^{-};\beta^{+}\right]^{T} \!\!\cdot\nabla \left( \!\left[  \begin{array}{c}
\!\! -P_{K}^{*} \!\!\\
\!\! P_{K}^{*} \!\!
\end{array}  \right] + \left[  \begin{array}{c}
\!\!\!\!\bar{H}_{2}\theta^{*}\!\!\!\!\\
\!\!\!\!-\bar{H}_{2}\theta^{*}\!\!\!\!
\end{array}  \right] \!-  M_{1 }\!\cdot \!\left[  \begin{array}{c}
\!\!1\! -  \!s\!\!\\
\!\!1\! -  \!s\!\!
\end{array}  \right] \!\right)\nonumber \\
& +\left[\gamma^{-};\gamma^{+}\right]^{T} \!\!\cdot\nabla \left(\! \left[ \begin{array}{c}
\!\!-P_{K}^{*}\!\!\\
\!\!P_{K}^{*}\!\!
\end{array} \right] \!-M_{1}\!\cdot\left[ \begin{array}{c}
\!\!s\!\!\\
\!\!s\!\!
\end{array} \right]\!\right)=0\label{eq:KKT_firstorder}\\
& \text{diag} \!\left( \!\left[ \mu^{-} ;\mu^{+} \right] \!\right) \!\cdot \left(\! \left[  \begin{array}{c}
\!\!\!\!-\bar{H}_{2}\left(\theta^{*} + c \right)\!\!\!\!\\
\!\!\!\!\bar{H}_{2}\left(\theta^{*} + c \right)\!\!\!\!
\end{array}  \right] \!- \!\left[ \begin{array}{c}
\!\!P_{K}^{\text{max}}\!\!\\
\!\!P_{K}^{\text{max}}\!\!
\end{array} \right] \!\right) =0\label{eq:DUALpowerflowlimit}\\
& \text{diag}\left(\!\left[\alpha^{-};\alpha^{+}\right]\!\right)\!\cdot\left(\!\left[\begin{array}{c}
\!\!-P_{G}^{*}\!\!\\
\!\!P_{G}^{*}\!\!
\end{array}\right]\!-\!\left[\begin{array}{c}
\!\!-P_{G}^{\text{min}}\!\!\\
\!\!P_{G}^{\text{max}}\!\!
\end{array}\right]\!\right)=0\label{eq:DUALgenlimit}\\
& \text{diag}\left(\!\left[\beta^{-};\beta^{+}\right]\!\right)\!\cdot\left(\! \left[  \begin{array}{c}
\!\!-P_{K}^{*}\!\!\\
\!\!P_{K}^{*}\!\!
\end{array}  \right] \!+ \!\left[  \begin{array}{c}
\!\!\!\!\bar{H}_{2}\theta^{*}\!\!\!\!\\
\!\!\!\!-\bar{H}_{2}\theta^{*}\!\!\!\!
\end{array}  \right] \!-  M_{1 }\!\cdot \!\left[  \begin{array}{c}
\!\!1\! - \! s\!\!\\
\!\!1\! - \! s\!\!
\end{array}  \right]\! \right)=0\label{eq:DUALpfLINEAR1}\\
& \text{diag}\left(\!\left[\gamma^{-};\gamma^{+}\right]\!\right)\!\cdot\left( \!\left[ \begin{array}{c}
\!\!-P_{K}^{*}\!\!\\
\vspace{-0.2cm}
\!\!P_{K}^{*}\!\!
\end{array} \right]\! -M_{1}\!\cdot\left[ \begin{array}{c}
\!\!s\!\!\\
\!\!s\!\!
\end{array} \right]\!\right)=0\label{eq:DUALpfLINEAR2}\\
& \left[\mu^{-};\mu^{+};\alpha^{-};\alpha^{+};\beta^{-};\beta^{-};\gamma^{-};\gamma^{+}\right]\geqslant0\label{eq:dualLimit}
\end{flalign}
where constraint \eqref{eq:KKT_firstorder} is the partial gradient
optimal condition, \eqref{eq:DUALpowerflowlimit}$-$\eqref{eq:DUALpfLINEAR2}
are the complementary slackness constraints, \eqref{eq:con_nodebalance}$-$\eqref{eq:con_GENlimit}
and \eqref{eq:con_linearPF} are the primal feasibility constraints,
and \eqref{eq:dualLimit} represents the dual feasibility constraints. 

Particularly, the complementary slackness constraints \eqref{eq:DUALpowerflowlimit}$-$\eqref{eq:DUALpfLINEAR2}
are nonlinear since they include product of continuous variables.
We then linearize them by introducing new binary variables $\delta_{\mu^{\pm}}$,
$\delta_{\alpha^{\pm}}$, $\delta_{\beta^{\pm}}$, and $\delta_{\gamma^{\pm}}$. For instance, constraint \eqref{eq:DUALpowerflowlimit} can be rewritten as  
\begin{equation}
\hspace{-0.2cm}\left\{ \begin{array}{lr}
\mu^{-}-M\cdot\delta_{\mu^{-}}\leqslant0\\
\bar{H}_{2}\left(\theta^{*}+c\right)+P_{K}^{\text{max}}\leqslant M\left(1-\delta_{\mu^{-}}\right)\\
\mu^{+}-M\cdot\delta_{\mu^{+}}\leqslant0\\
-\bar{H}_{2}\left(\theta^{*}+c\right)+P_{K}^{\text{max}}\leqslant M\left(1-\delta_{\mu^{+}}\right)
\end{array}\right.\label{eq:MIP_PFlimit}\\
\end{equation}
where $M$ is a large positive number. Constraints \eqref{eq:DUALgenlimit}$-$\eqref{eq:DUALpfLINEAR2} can be linearized using the same method. Particularly, for the linearized forms of constraints \eqref{eq:DUALpfLINEAR1} and \eqref{eq:DUALpfLINEAR2}, $M_{1}$ and
$M$ are different values and $M_{1}\ll M$.

Using the approximate relaxation for the various constraints as detailed above, we obtain the following equivalent single-stage mixed-integer linear problem with the objective 
\begin{equation}
\text{max} \hspace{0.2cm} P_{Kl}-\zeta\underset{^{n\in\mathcal{L}}}{\sum}u_{n}  \label{eq:Mod_Obj}
\end{equation}
 subject to \eqref{eq:con_tripline}, \eqref{eq:con_loadshift}, \eqref{eq:con_nodebalance}, \eqref{eq:con_powerflow}$-$\eqref{eq:con_linearPF}, \eqref{eq:con_linearN1}, \eqref{eq:KKT_firstorder}, \eqref{eq:dualLimit}, \eqref{eq:MIP_PFlimit}, and the linearized forms of constraints \eqref{eq:DUALgenlimit}$-$\eqref{eq:DUALpfLINEAR2}. Note that since no real-time data is required in the above optimization problem, the attacker can solve this step offline to determine the switching attack line to trip and the attack vector $c$.
\subsection{\label{sub:Initial-attack-vector}Step 2: Determine Initial Attack Vector}
 In this step, we determine the attack vector at events $A_{1}$. As stated earlier, $c^{0}$, the attack vector at $A_{1}$, is chosen to ensure that the resulting load shifts lead to the optimal dispatch solved in Step 1, \textit{i.e.,} $P_{G}^{*}$. To this end, we use a two-stage optimization problem similar to Step 1 to determine $c^{0}$. Note that since the switching attack line and attack sub-graph are both determined in Step 1, the dependency matrix between power injection and voltage angle, \textit{i.e.,} $H_{1}$ for the physical topology $\mathcal{G}$ at $A_{1}$ is known to the attacker, and the cyber loads are given by \eqref{eq:falseload_A1}. 

As stated in Section \ref{sec:Gen-attack-model}, the attacker only has access to the measurements inside $\mathcal{S}$. Thus, the attacker cannot directly obtain the whole system physical states $\theta_{0}$. However, assuming $\mathcal{B}_{\mathcal{S}}$ and $\mathcal{B}_{N\mathcal{S}}$ represents the set of buses inside and outside $\mathcal{S}$, respectively. The vector of cyber loads resulting from an unobservable attack satisfies the following relationship:
\begin{equation}
\bar{P}_{D}=
\left[ \begin{array}{lr}
P_{D}^{\mathcal{B}_{\mathcal{S}}}+H_{1}^{\mathcal{B}_{\mathcal{S}}}\theta_{0}^{\mathcal{B}_{\mathcal{S}}}-\bar{H_{1}}^{\mathcal{B}_{\mathcal{S}}}\left(\theta_{0}^{\mathcal{B}_{\mathcal{S}}}+c^{0}\right)  \\
P_{D}^{\mathcal{B}_{N\mathcal{S}}}
\end{array}\right] \label{eq:cyberload1}
\end{equation}
where $P_{D}^{\mathcal{B}_{\mathcal{S}}}$ is the vector of physical loads for all buses inside $\mathcal{S}$, $P_{D}^{\mathcal{B}_{N\mathcal{S}}}$ is that for all buses outside $\mathcal{S}$,  $H_{1}^{\mathcal{B}_{\mathcal{S}}}$ and $\bar{H}_{1}^{\mathcal{B}_{\mathcal{S}}}$ represents the sub-matrices of $H_{1}$ and $\bar{H}_{1}$ for the set of buses inside $\mathcal{S}$, respectively. For the physical system states $\theta_{0}^{\mathcal{B}_{\mathcal{S}}}$, attacker uses the estimated states $\hat{\theta}_{0}^{\mathcal{B}_{\mathcal{S}}}$, to compute \eqref{eq:cyberload1}. 

The two-stage optimization can be written as follows:
\begin{flalign}
\text{min}\;\; & \left\Vert c_{\mathcal{L}}\right\Vert _{0}\label{eq:OBJ_minSubgraph-1}\\
\text{s.t.}\;\; & -\tau P_{D}^{\mathcal{B}_{\mathcal{S}}}\leqslant\bar{H}_{1}^{\mathcal{B}_{\mathcal{S}}}\left(\hat{\theta}_{0}^{\mathcal{B}_{\mathcal{S}}}+c^{0}\right)-H_{1}^{\mathcal{B}_{\mathcal{S}}}\hat{\theta}_{0}^{\mathcal{B}_{\mathcal{S}}}\leqslant\hspace{-0.05cm} \tau P_{D}^{\mathcal{B}_{\mathcal{S}}}\label{eq:con_loadshift_H-1}\\
& \left\Vert c_{\mathcal{L}}\right\Vert _{0}\leqslant N_{0}
\label{eq:con_linearizedN1-1}\\
& \left\{ \bar{\theta}^{*} ,  P_{G}^{*} \right\}  =arg \left\{   \underset{\bar{\theta},P_{G}}{\text{min}}\:\underset{g=1}{\overset{n_{g}}{\sum}}  C_{g} \left(P_{Gg}\right) \right\} \label{eq:OBJ_MINCOST-1-1}\\
& \text{s.t.}\hspace{0.1cm}A_{GN}\cdot P_{G}-\bar{H}_{1}\cdot\bar{\theta} \nonumber\\
& \hspace{0.5cm}=\left[\hspace{-0.1cm} \begin{array}{lr}
P_{D}^{\mathcal{B}_{\mathcal{S}}}+H_{1}^{\mathcal{B}_{\mathcal{S}}}\hat{\theta_{0}}^{\mathcal{B}_{\mathcal{S}}}-\bar{H_{1}}^{\hspace{-0.1cm}\mathcal{B}_{\mathcal{S}}}\hspace{-0.1cm}\left(\hspace{-0.1cm}\hat{\theta_{0}}^{\mathcal{B}_{\mathcal{S}}}+c^{0}\hspace{-0.1cm}\right)  \\
P_{D}^{\mathcal{B}_{N\mathcal{S}}}
\end{array}\hspace{-0.1cm}\right] \hspace{0.1cm} (\lambda) \label{eq:con_nodebalance-1-1}\\  
& \begin{array}{cc}
\hspace{0.2cm}-  P_{K}^{\text{max}} \leqslant \bar{H}_{2}\bar{\theta} \leqslant  P_{K}^{\text{max}} & (\mu^{-},\mu^{+})\end{array}\label{eq:con_powerflow-1-1}\\
& \begin{array}{cc}
\ \hspace{0.1cm} P_{G}=P_{G1}^{*} & (\alpha)\end{array}\label{eq:con_GENlimit-1-1}
\end{flalign}
where $P_{G1}^{*}$ is $n_{g}\times1$ optimal generation
vector solved in Step 1; $\alpha$ is $n_{g}\times1$ dual variable
vector of constraints \eqref{eq:con_GENlimit-1-1}. The objective
\eqref{eq:OBJ_minSubgraph-1}
is to minimize the $l_{0}-$norm of the attack vector. Constraint
\eqref{eq:OBJ_minSubgraph-1} represents load shift limitation. Constraints
\eqref{eq:OBJ_MINCOST-1-1}$-$\eqref{eq:con_GENlimit-1-1} represent
the second stage DCOPF problem, which guarantees that the attack vector
selected in the first stage leads to the optimal dispatch $P_{G1}^{*}$.
The $l_{0}-$norm constraint can be relaxed to a linearized $l_{1}-$norm constraint as \eqref{eq:NORM1}. The objective can be represented as 
 $\underset{^{n\in\mathcal{L}}}{\sum}u_{n}$. 
 This problem can then be converted to a single stage optimization
problem using methods similar to those as in detailed Step 1.  

%\vspace{-0.4cm}
\subsection{\label{sub:Implementation}Implementation}

The method to construct an unobservable AC attack with a DC attack vector has been introduced in \cite{Liang2015} for FDI attacks without topology alteration. In this paper, we focus on constructing AC unobservable cyber-physcial attacks. The procedure is as follows:

\begin{enumerate}
	\item Solve the Step 1 optimization offline to obtain the switching attack line $t$ and the attack vector $c$.
	\item Identify the attack sub-graph $\mathcal{S}$ with $c$ and line $t$.
	\item Launch the physical attack on the switching attack line.
	\item Perform local SE inside $\mathcal{S}$ with slack bus chosen as one arbitrary load bus in $\mathcal{S}$ to obtain $\hat{\theta}_{0}^{\mathcal{B}_{\mathcal{S}}}$; 
	\item Solve the Step 2 optimization problem to obtain $c^{0}$; 
	\item For all load buses $m$ inside $\mathcal{S}$, set $\bar{\theta}_{0m}=\hat{\theta}_{0m}+c^{0}_{m}$; 
	\item For all non-load buses, since the net injections are not changed, the nodal balance equations for each non-load buses are
	\begin{flalign}
	\hspace{-0.6cm} A_{GN}^{m} P_{G}\negthinspace-\negthinspace V_{m}\negthickspace\negthickspace\underset{j\in\bar{\mathcal{B}}r_{m}}{\sum}\negthickspace  V_{j}(G_{mj}\text{cos}\theta_{mj}+B_{mj}\text{sin}\theta_{mj})\negthickspace=\negthickspace 0\\
	\hspace{-0.6cm}	A_{GN}^{m} Q_{G}\negthinspace-\negthinspace V_{m}\negthickspace\negthickspace\underset{j\in\bar{\mathcal{B}}r_{m}}{\sum}\negthickspace V_{j}(G_{mj}\text{cos}\theta_{mj}-B_{mj}\text{sin}\theta_{mj})\negthickspace=\negthickspace 0 
	\end{flalign}
	where $A_{GN}^{m}$ represents the $m^{\text{th}}$ row of $A_{GN}$, $Q_{G}$ represents the reactive power generation vector, $G_{mj}+iB_{mj}$ is the $(m,j)^{\text{th}}$ entry of the bus admittance matrix, and $\theta_{mj}=\theta_{m}-\theta{j}$ is the voltage angle difference between bus $m$ and $j$, $\bar{\mathcal{B}}r_{m}$ is the set of branches connecting to bus $m$ for $\bar{\mathcal{G}}$. These equations can be solved iteratively with Newton-Raphson method.
	\item After updating the cyber states for the non-load buses, using equation \eqref{eq: ModMeasureStra1} to calculate the AC attack for $A_{1}$.    
	\item Repeat Steps 4)$-$7) (without solving Step 2 optimization) to construct AC attacks with $c$ for $A_{2,...,T}$.
\end{enumerate}

%The temporal nature of system operation within attack is shown in
%Fig. \ref{fig:Temporal-nature-of-1}.

%\vspace{-0.3cm}
\section{\label{sec:Simulation}Numerical Results}

In this section, we test
the effect of attacks designed with the two-step attack
strategy for a nonlinear system model. The test system is
the IEEE 24-bus reliable test system (RTS). We assume: (i) the system is
operating under optimal power flow; and (ii) the loads of the system are
constant and are equivalent to the historic load data that is assumed to be known to the attacker. To model realistic power
systems, we assume that there are congestions prior to the
attack and the attacker chooses one congested line as target to maximize
power flow. We use MATPOWER to run AC power flow and AC OPF. The optimization problem is solved with CPLEX. 

%\vspace{-0.4cm}
\subsection{Solution for the attack designed with the
	attack strategy}

\begin{figure}[tbh]
	\centering\includegraphics[scale=1.2]{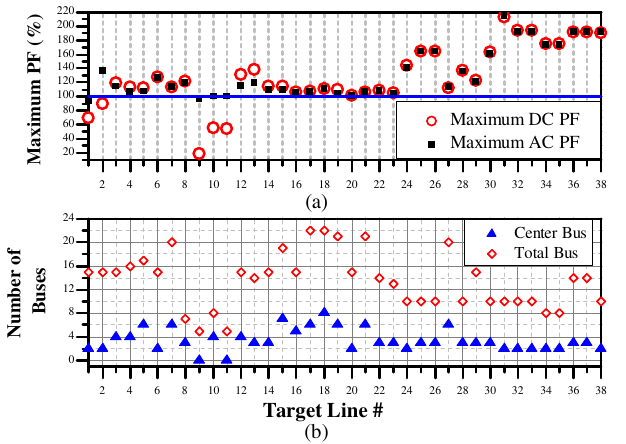}\vspace{-0.1cm}
	
	{\footnotesize{} (a) Maximum DC \& AC PF (b) Number of Center Buses \& Total Buses inside Attack Sub-graph v.s. Target Line.} %\vspace{-0.2cm}}

	\protect\caption{Summary of all 38 target lines under with $\tau=10\%$ and $N_{1}=0.06$.\label{fig:Summary-of-all}}
	\vspace{-0.2cm}
\end{figure}

The solution of the unobservable
topology attack determined by the two-step attack strategy is tested in this subsection. In order to understand the worst-case effect of attacks, we assume there is a line congested prior to the physical attack. This is achieved in simulation by reducing the line
rating to 95\% of the base case power flow (apparent power) to create congestion. We exhaustively
test all 38 lines as targets in the system and let $\zeta$, the weight for the $l_{1}-$norm term in the objective in \eqref{eq:Mod_Obj}, be 1\% of the original power flow on each target line.
Fig. \ref{fig:Summary-of-all} illustrates the maximal power flow (PF) and attack size (\# of buses in sub-graph) for load shift bounds $\tau=10\%$, total lines to physically attack $N_{T}=1$, and the $l_{1}$-norm constraint
$N_{1}=0.06$. The plot in Fig. \ref{fig:Summary-of-all}(a) indicate the flow attack end of system event $S_{1}$ using attack vector from event $A_{1}$. In Fig. \ref{fig:Summary-of-all}(a), we compare the physical power
flow (apparent power, we denote it as AC PF) in each line to the power flow solved in linear model (we denote it as DC PF). In Fig. \ref{fig:Summary-of-all}(b), we plot the number of center buses, \textit{i.e.}, $l_{0}-$norm of the attack vector, and the total number of buses inside the attack sub-graph for each target line. 

%\vspace{-0.2cm}	

From Fig. \ref{fig:Summary-of-all}(a) we can observe that the attack
vector determined by the two-stage optimization problem cause
overflows in 33 target lines in linear model, \textit{i.e.}, 86.84\% of the attacks are successful. For all such successful attacks, using the attack vector to construct an attack in the nonlinear model, in Fig. \ref{fig:Summary-of-all}(a), the AC PF in each line tracks DC PF solved with the attack strategy. In particular, 2 cases with target lines 9 and 11, respectively, have no center buses, \textit{i.e.}, for these lines the state-preserving attacks introduced in \cite{Jzhang2015} suffice. In Fig. \ref{fig:Summary-of-all}(b), we can observe that 72.73\% of the successful attacks can be launched inside a sub-graph with less than 16 total buses. 
  
%\vspace{-0.1cm}
\begin{figure}[tbh]
	\centering\includegraphics[scale=0.65]{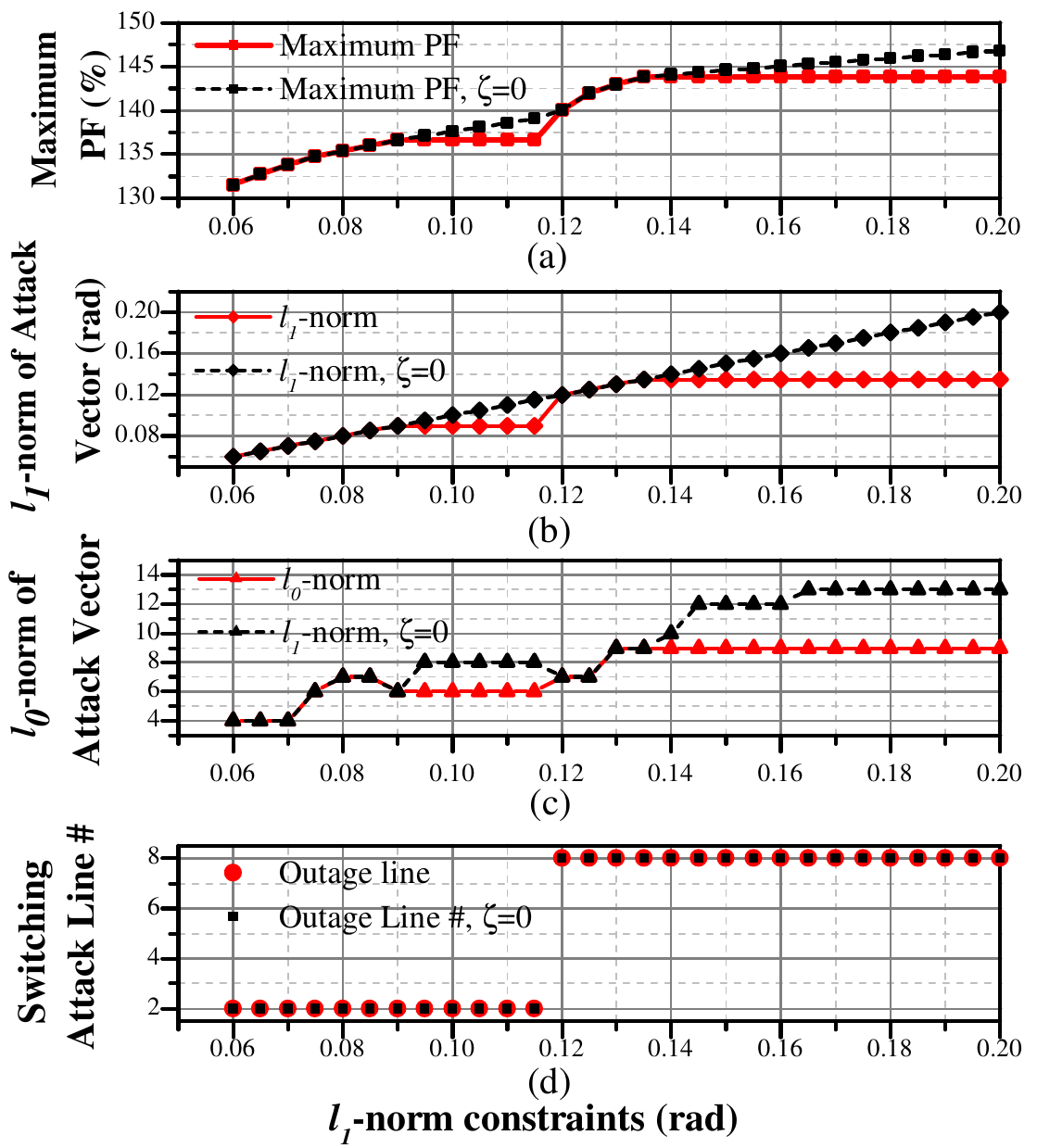}%\vspace{-0.1cm}
	
	{\footnotesize{} (a) Maximum PF (b) $l_{1}-$norm of attack vector (c) $l_{0}-$norm of attack vector (d) the switching attack line v.s. the conditional $l_{1}-$norm constraints.\vspace{-0.2cm}}

	\protect\caption{Target Line 12 (Connecting Bus 8 - Bus 9) with $\tau=10\%$.
		\label{fig:Target-line-=00002312}}
	\vspace{-0.1cm}
\end{figure} 
In Fig. \ref{fig:Target-line-=00002312}, we illustrate the effect of the $l_{1}-$norm constraint on the maximal power flow (Fig. \ref{fig:Target-line-=00002312}(a)), the $l_{1}-$ and the $l_{0}-$norms of the attack vector (Figs. \ref{fig:Target-line-=00002312}(b) and \ref{fig:Target-line-=00002312}(c), respectively), and the switching attack line (Fig. \ref{fig:Target-line-=00002312}(d)) for target line 12 solved with Step 1 optimization. In each sub-figure, we illustrate the two solutions: one with $\zeta$ (red) and the other without $\zeta$ (black) in the objective function. From Figs. \ref{fig:Target-line-=00002312} (a)$-$(c), we can see that for the solutions without  $l_{1}-$norm in objective (\textit{i.e.}, $\zeta=0$) as the $l_{1}-$norm constraint is relaxed, the maximal target line power flow as well as the $l_{1}-$ and $l_{0}-$norms of attack vector also increase. In contrast, for plots with the $l_{1}-$norm in the objective, the $l_{1}-$norm ensures that the vector with the smallest number of center buses is chosen. This in turn implies that when the $l_{1}-$norm in the objective is tight, the resulting power flow may be smaller than that obtained without such a constraint. These differences are illustrated in Figs. \ref{fig:Target-line-=00002312}(a)$-$(c). In Fig. \ref{fig:Target-line-=00002312}(d), we demonstrate that
the switching attack line chosen by the optimization problem changes from line 2 to line 8 as the $l_{1}-$norm constraint is relaxed. In general, tripping line 8 requires a large load shift, and thus, is only possible for larger $l_{1}-$norm constraint as then the cyber load changes can be distributed over a larger number of load buses in the sub-graph. %The maximal target line flow can result from multiple attack vectors and the one related to the smallest attack sub-graph is favored.  
%However, since there are two conflict terms, \textit{i.e.}, maximize the target line power flow while minimize the $l_{1}$-norm of the attack vector in the objective, the optimal solution may not guarantee the maximal power flow on the line but the chosen power flow is close enough to the maximal values with the smallest attack sub-graph. 
%In Fig. \ref{fig:Target-line-=00002312}(b), the switching attack line chosen for the target line 12 overlaps for cases with or without the objective that minimize $l_{1}$-norm of the attack vector. As the $l_{1}$-norm constraint increase, the switching attack line also changes from line 2 to line 8. This is because at this constraint, load shifts are allowed to redistribute to more buses, and thus, the outage line 8 that can lead to higher overflow while requiring larger load shifts will be chosen.
%The solution is a trade-off between the maximal power flow and minimal attack sub-graph.  
%that maximize the target line power flow while minimize the $l_{1}$-norm of the attack vector. 
   
%the $l_{1}-$norm in the objective of the Step 1 optimization problem ensures that the vector with the smallest number of center buses is chosen. This in turn implies that when the $l_{1}-$norm in the objective is tight, the resulting power flow may be smaller than that obtained without such a constraint. It is these differences that are illustrated in Fig. \ref{fig:Summary-of-all} (a).

%\vspace{-0.4cm}
\subsection{Consequences of the attack in the nonlinear model}

%\begin{figure}[tbh]
%	\centering\includegraphics[trim=0cm 0.1cm 0cm 0.1cm, clip=true,scale=0.32]{24rts_graph_N}\protect\caption{Sub-graph of Attack Case When Line 12 Congested and Line 2 Has
%		A Physical Outage within Unobservable Topology Attack. \label{fig:Subgraph-of-attack} }
%	\vspace{-0.5cm}
%\end{figure}

In this subsection, we select a typical case to demonstrate the
consequence of the unobservable state-and-topology cyber-physical attack determined by the attack strategy in the nonlinear system model. In this case, the target line is line 12 with $\tau=10\%$, $N_{T}=1$, and $N_{1}=0.06$. Under this condition, the switching attack line
is line 2. 

%%The minimum sub-graph
%%for this case is shown in Fig. \ref{fig:Subgraph-of-attack}. In this figure, the attack sub-graph is demonstrated in red and blue. The red part is determined by the center buses 4, 7, 8, and 10, while the blue part is to complete the path connecting the switching attack buses. 

\begin{figure}[tbh]
	\centering\includegraphics[scale=1]{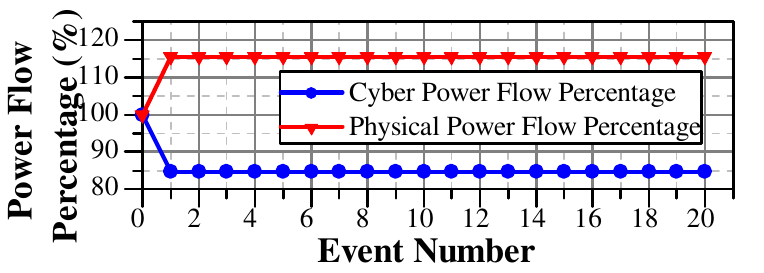}\vspace{-0.35cm}\protect\caption{Power Flow Variation on Line 12 During 20 System Events.\label{fig:Power-flow-on12}}
	\vspace{-0.1cm}
\end{figure}

For the chosen target line, after launching the physical attack at $A_{0}$ and injecting the initial cyber attack constructed with $c^{0}$ at $A_{1}$, the active power generation dispatch
for generators at bus 7 and 13 change from $215.69$ MW and $230.96$ MW to $200.69$ MW and $245.67$ MW, respectively
(the dispatch of other generators remain unchanged). In the following events, as the attacker continues to inject the AC attacks constructed with attack vector $c$ (determined by Step 1 optimization), the active power generation for these two set of generators are maintained at these values. Fig. \ref{fig:Power-flow-on12} demonstrates the cyber and physical
power flow variation during $20$ system events.  
From Fig. \ref{fig:Power-flow-on12},
we can observe that once the active power generation dispatch changes to the optimal dispatch and remains unchanged in the subsequent system events, the physical
overflow in the target line will be maintained by injecting the AC attack constructed with attack vector $c$. The heat accumulation may eventually
cause this line to overheat and then trip offline all the while remaining unobservable to the control center. 
	\vspace{-0.1cm}

%\begin{figure}[tbh]
%	\centering\includegraphics[trim=0cm 0cm 0cm 0.2cm, clip=true, scale=0.8]{Gen_PF_variation}\protect\caption{Active Power Output Variation When Line 2 Has Physical Outage and
%		12 Was Congested Prior to Unobservable Topology Attack.\label{fig:Active-power-output12}}
%	\vspace{-0.3cm}
%\end{figure}
%\vspace{-0.5cm}

%\begin{figure}[tbh]
%	\centering\includegraphics[scale=1]{loadshift12_n150ld10}\vspace{-0.3cm}\protect\caption{Load Shift Percentage for Each Bus of the Test System.\label{fig:Load-shift-percentage12}}
%	%\vspace{-0.3cm}
%\end{figure}

%In Fig. \ref{fig:Load-shift-percentage12}, we compare the load
%shifts caused by the attack determined by the two-step optimization problem in linear model and the
%actual load shifts caused by the AC attack constructed with the DC attack vector in the nonlinear model. We find that for the attack in nonlinear system model, the load shifts
%caused by the attack are less than 10\% and are comparable to those solved in linear model.

%A natural question that can be asked is that relative to the optimization problem which models the attack in the linear system, are the load shifts caused by the attack in the nonlinear system model still within the same bound. 
%In Fig. \ref{fig:Load-shift-percentage12}, we demonstrate the load shifts caused by the attack in the linear system model (denote as DC) and that in the nonlinear system model (denote as AC) for this case. We find that for the attack in the nonlinear system model, the load shifts
%caused by the attack are still within 10\% load shift bounds. 
We compare the load shifts caused by the attacks in both the linear and nonlinear system models and find that the load shifts in the nonlinear system model track those in the linear system model for most of the successful attacks. The only two exceptions are for lines 13 (\textit{i.e.}, 20\% load shift on a bus) and 23 (\textit{i.e.}, 15\% load shift on a bus).

For the successful attacks with other target lines, we observe similar attack consequences in the nonlinear model. 

%In addition, we compare the load
%shifts calculated by the optimization problem in linear model and the
%actual load shifts in the nonlinear model. We find that for the attack in nonlinear system model, the load shifts
%caused by the attack are less than 10\% and are comparable to those solved in linear model. 

%\vspace{-0.3cm}
\section{\label{sec:conclusion}Concluding Remarks}

In this paper, we have introduced a class of unobservable topology attacks in which
both topology data and states for a sub-graph of the network are
changed by an attacker. We have proposed a two-step attack strategy to maximize the power flow on a target line subject to constraints on limited size of attack sub-graph and limited load shifts. We have shown that attacks designed
with the proposed two-step attack strategy can cause physical line overloads in the IEEE 24-bus RTS even when the attack is subject to bounds on changes in load, for both linear and nonlinear models. The proportion of the successful attacks in the nonlinear system model is 86.84\%, which shows the vulnerability of the system to such attacks.

A potential countermeasure is to use historical data to forecast and predict expected generation dispatch. The cyber load patterns created by the attack will in general be different from the normal load shift patterns that lead to the same dispatch plan. Thus, such forecasting can lead to detection of anomalies in both loads and dispatch.  

An important extension to study is to understand the impact of attacks when the attacker has access to topology and generation data only for a sub-network. While our attack model restricts data changes to a sub-graph, it still requires the attacker to have knowledge of the complete system topology and generation data. Yet another avenue is to study the worst-case attacks that trip multiple switching attack lines and maximize power flow on multiple target lines. 

\bibliographystyle{IEEEtran}
\bibliography{DistributedAttacks}
%\begin{biography}[{\includegraphics[width=1in,height=1.25in,clip,keepaspectratio]{mshell}}]{Michael Shell}

%\end{IEEEbiography}
%% insert where needed to balance the two columns on the last page with
%%\newpage


%\vfill


\end{document}